\documentclass[aps,prl,superscriptaddress,twocolumn,longbibliography]{revtex4-1}
\usepackage{bbm}
\usepackage{graphicx}
\usepackage{dcolumn}
\usepackage{bm}
\usepackage{subfigure}
\usepackage{amsmath}
\usepackage{feynmf}
\usepackage{hyperref}

\usepackage{attachfile}

\usepackage{times}

\def\be{\begin{equation}} \def\ee{\end{equation}}
\def\bea{\begin{eqnarray}} \def\eea{\end{eqnarray}}

\def\bq{{\bf q}}

\def\bk{{\bf k}}

\def\bK{{\bf K}}
\def\be{{\bf e}}
\def\bd{{\bf d}}

\def\rw{\rightarrow}

\begin{document}

\title{Precise determination of critical points of topological phase transitions via shift current\\
in two-dimensional inversion asymmetric insulators}

\author{Zhongbo Yan}
\email{yanzhb5@mail.sysu.edu.cn}
\affiliation{ School of Physics, Sun Yat-Sen University, Guangzhou 510275, China}

\begin{abstract}
The precise determination of critical point is the basis to extract various critical properties of phase transitions.
We identify that for two-dimensional inversion asymmetric insulators, with and without time-reversal symmetry,
when topological phase transitions take place,  all nonvanishing components of band-edge shift current tensor will reverse their signs
in a singular way, regardless of what realistic value the temperature takes.
This remarkable sign-reversal behavior of band-edge shift current tensor thus can be applied to
determine the critical points of various topological phase transitions precisely,
even for temperature-driven ones. We suggest concrete materials to test our predictions.



\end{abstract}

\maketitle

Since the first successful exfoliation of graphene\cite{Novoselov2004}, the family of two-dimensional (2D) materials
has grown extremely fast over the last decade, ranging from insulators, metals to superconductors\cite{Miro2014,Novoselov2016review,Wang2012review}.
Owing to their atomically thin structures, 2D materials have demonstrated various novel phases,
as well as fascinating
electronic, optical and mechanical properties that do not exist in their bulk counterparts\cite{Wang2012review,Geim2013,Butler2013,Xia2014,Fiori2014}. Another remarkable common feature of 2D materials is that
their atomically thin structures also provide exceptional flexibility to tailor 
their bulk properties\cite{Castro2007,Gui2008,Mak2009,Ni2012,Drummond2012}, suggesting that
2D materials are ideally suited for an in-depth investigation of various competing phases,
as well as their transitions and critical properties\cite{Yu2015,Li2015}.
The emerged 2D materials also open up a new door for topological phases and related physics\cite{Yan2012review,Ando2013review,Ren20106review}.
The most notable example is the demonstration of a quantum spin Hall insulating phase in graphene with
intrinsic spin-orbit coupling\cite{kane2005a, kane2005b}.  This conceptional breakthrough reveals
 that seemingly featureless band insulators actually have very rich physics and need to be further
classified  according to
their underlying band topology\cite{thouless1982,kane2005b,fu2007b,fu2007a,
moore2007,qi2008,schnyder2008,kitaev2009,ryu2010,wang2010b,wang2012a,hasan2010,qi2011,Bansil2016}

The existence of distinct topological phases in insulators raises a natural question: what kind of new critical properties may emerge
at the critical points of continuous topological phase transitions (TPTs) between distinct insulating phases.
While the critical properties of TPTs have been actively studied in
theory\cite{Ostrovsky2011,Goswami2011,Gulden2016transition,Bagrets2016transition,
Roy2016TPT,Chen2017TPT,Wang2017transition,Goswami2017,Zeng2017TPT,Griffith2018}, thus far they have been
largely  unexplored in experiments, mainly owing to the lack of experimental methods that are able to
determine the critical points of TPTs precisely. As is known, the critical points of conventional continuous phase transitions
can simply be determined in experiments by the observation of singular behavior in physical quantities.
For TPTs, as the name suggests, the most dramatic change are topological invariants, which,
however, are not physical observables.
Generally, the topological invariant of an insulating phase is revealed by
measurements of quantized linear-response transport coefficients\cite{klitzing1980,tsui1982,konig2007,chang2013experimental},
however, owing to finite temperature effect and various scattering effects,
the quantization of linear-response transport coefficients breaks down
when the system gets closed to the critical point at which the band gap vanishes, indicating the absence of
any singular signature in these observables at the critical point\cite{He2017chiral}.

In this work, we show that a measurement of shift current (a nonlinear optical effect\cite{Baltz1981bpve,Fridkin2001bpve,Sipe2000bpve})
can determine the critical points of TPTs in 2D inversion
asymmetric insulators precisely, owing to that the band-edge shift current tensor
will display a singular sign-reversal behavior
across the critical points. As is known,
only two of the ten symmetry classes, class AII (with time-reversal symmetry (TRS)) and class A (without TRS),
can host topological insulating phases in 2D\cite{schnyder2008,kitaev2009}.The former is classified by a $Z_{2}$ number,
while the latter is classified by an integer, the Chern number\cite{schnyder2008,kitaev2009}. Remarkably, we demonstrate compactly
that in the absence of inversion symmetry (a prerequisite
for the presence of shift current\cite{Baltz1981bpve,Fridkin2001bpve,Sipe2000bpve}), all nonvanishing components of band-edge shift current
tensor will reverse their signs in a singular way across the TPTs allowed by these two symmetry classes,
regardless of what realistic value the temperature takes,
indicating the wide applicability of this approach in 2D. Noteworthily,  while in the absence of inversion symmetry
certain component of shift current tensor was also found to reverse its sign when
a normal insulator is transited to a topological insulator in 3D\cite{Tan2016bpve}, the transition itself is
in general indirect owing to the existence of an intermediate gapless phase\cite{murakami2007phase,murakami2008}.

{\it General theory.---} Near the critical point of a continuous TPT,
the relevant physics are faithfully described by the low-energy Hamiltonian around the band edge
with the smallest energy gap. For generality, we consider the low-energy Hamiltonian takes the from $H=\sum_{q;\alpha}\oplus \Psi_{\alpha;\bq}^{\dag}H_{\alpha}(\bq)\Psi_{\alpha;\bq}$ with $\Psi_{\alpha}=(c_{\alpha;\bq; 1},...,c_{\alpha;\bq; n})^{T}$ an
$n$-component spinor,
$\alpha$ labeling the $\alpha$-th band edge,  and $\bq=(q_{x}, q_{y})$ the momentum relative
to the band edge. Different $H_{\alpha}(\bq)$ are related by symmetry, so that they are degenerate in energy
and their energy gaps get closed and reopened at the same time.

For 2D inversion asymmetric insulators without TRS, the low-energy Hamiltonian is in general a rank-$2$ matrix.
The presence of TRS in general requires it  to be a rank-$4$ matrix,
but if spin conserves, it can be reduced as the direct sum of two rank-$2$ matrices. For simplicity,
we first confine ourselves to the spin conserving case if TRS is respected
 and address the spin non-conserving case later. Then to second order in momentum, the general form of $H_{\alpha}(\bq)$  is
$H_{\alpha}(\bq)=\bd_{\alpha}(\bq)\cdot \mathbf{\tau}+\epsilon_{\alpha}(\bq)\mathbf{I}$, with
\begin{eqnarray}
d_{\alpha;x}(\bq)&=&\Delta_{\alpha;x}+v_{\alpha;x}q_{x}+A_{\alpha;ij}q_{i}q_{j}, \nonumber\\
d_{\alpha;y}(\bq)&=&\Delta_{\alpha;y}+v_{\alpha;y}q_{y}+B_{\alpha;ij}q_{i}q_{j},\nonumber\\
d_{\alpha;z}(\bq)&=&m_{\alpha}+\lambda_{\alpha;x}q_{x}+\lambda_{\alpha;y}q_{y}+C_{\alpha;ij}q_{i}q_{j},\label{general}
\end{eqnarray}
$\mathbf{\tau}=(\tau_{x},\tau_{y},\tau_{z})$  the Pauli matrices, and $\mathbf{I}$ the rank-2 unit matrix.
We will set $\epsilon_{\alpha}(\bq)=0$ below, since it is irrelevant to the physics we will discuss.
The parameters $\{\Delta_{\alpha;x,y}, m_{\alpha}, v_{\alpha;x,y},
\lambda_{\alpha;x,y}, A_{\alpha;ij}, B_{\alpha;ij}, C_{\alpha;ij}\}$ are all momentum-independent,  and $X_{ij}q_{i}q_{j}$ with $X=\{A, B, C\}$ is
a shorthand notation of $X_{xx}q_{x}^{2}+X_{yy}q_{y}^{2}+X_{xy}q_{x}q_{y}$.
To ensure that the Hamiltonian correctly describes the band edge,
all linear momentum terms in the energy spectra ($E_{\alpha}^{\pm}=\pm d_{\alpha}$, with $d_{\alpha}\equiv\sqrt{d_{\alpha;x}^{2}+d_{\alpha;y}^{2}+d_{\alpha;z}^{2}}$)
must vanish, which puts  the following two constraints
on the above parameters,
\begin{eqnarray}
\Delta_{\alpha;x}v_{\alpha;x}=-m_{\alpha}\lambda_{\alpha;x},\quad \Delta_{\alpha;y}v_{\alpha;y}=-m_{\alpha}\lambda_{\alpha;y}.\label{constraint}
\end{eqnarray}

Topological properties of $H_{\alpha}(\bq)$ are characterized by the Chern number\cite{Xiao2010Berry} (note if TRS is preserved,
$H_{\alpha}$ has a TRS partner with opposite Chern number, and the total Hamiltonian $H$ is characterized by a $Z_{2}$ number).
Neglect all quadratic momentum terms in $H_{\alpha}(\bq)$, a short calculation reveals that the Chern number $C_{\alpha}=\frac{1}{2}\text{sgn}(v_{\alpha;x}v_{\alpha;y}m_{\alpha})$,
indicating that $H_{\alpha}(\bq)$ will undergo a TPT when
the mass term $m_{\alpha}$ changes sign. As in general the coefficients
of the terms containing momentum  will keep their signs
when $m_{\alpha}$ is varied across the critical point,
the constraints in Eq.(\ref{constraint}) indicate that
as long as $\Delta_{\alpha;x}$ and  $\Delta_{\alpha;y}$ do not identically equal zero ($\Delta_{\alpha}\equiv0$ implies
$\lambda_{\alpha}\equiv0$), they will change their signs with $m_{\alpha}$ at the same time.

Let us now investigate the shift current, which is a second-order optical effect with the induced direct current proportional
to the square of optical field\cite{Baltz1981bpve,Fridkin2001bpve,Sipe2000bpve,Cook2017bpve,Young2012bpve1,
Young2012bpve2,Morimoto2016nloe,Kim2017TIbpve,Rangel2017shift,Fregoso2017shift,Wang2017shift,Hiroaki2017shift}), i.e., $J_{a}=\sigma^{abb}(\omega)\mathcal{E}_{b}(\omega)\mathcal{E}_{b}(-\omega)$, where
$\sigma^{abb}$ represents the shift current tensor, and $\mathcal{E}_{b}$ denotes the optical field.
For the two-band Hamiltonian $H_{\alpha}(\bq)$, the shift current tensor is simply
determined by the following formula\cite{Sipe2000bpve,Cook2017bpve, shiftcurrent}
\begin{eqnarray}
\sigma_{\alpha}^{abb}(\omega)&=&\frac{4\pi e^{3}}{\hbar^{4}\omega^{3}}\int\frac{d^{2}q}{(2\pi)^{2}}
[F_{1\alpha}^{abb}+F_{2\alpha}^{abb}] f_{\alpha}^{-+}
\delta(\hbar\omega-2d_{\alpha}),\quad\label{formula}
\end{eqnarray}
where $a, b=\{x,y\}$, $f_{\alpha}^{-+}=f(-d_{\alpha})-f(d_{\alpha})$, with $f(d_{\alpha})=1/\{1+\text{exp}[(d_{\alpha}-\mu)/k_{\rm B}T]\}$
the Fermi-Dirac distribution function ($\mu$ the chemical potential, $T$ the temperature, and $k_{B}$ the Boltzmann constant).
Expressions for the two integrands in bracket are $F_{1\alpha}^{abb}=\bd_{\alpha}\cdot(\partial_{ab}\bd_{\alpha}\times\partial_{b}\bd_{\alpha})$ and
$F_{2\alpha}^{abb}=(\hbar\omega)^{2}\Omega_{\alpha;ab}\partial_{b}d_{\alpha}/2$, with $\Omega_{\alpha;ab}=-\bd_{\alpha}\cdot(\partial_{a}\bd_{\alpha}\times\partial_{b}\bd_{\alpha})/2d_{\alpha}^{3}$
the Berry curvature of the valence band\cite{Xiao2010Berry}.
Because $\Omega_{\alpha;ab}$ vanishes identically when $a=b$, it is readily seen
that $F_{2\alpha}^{xxx}$ and $F_{2\alpha}^{yyy}$ also vanish identically.
For $F_{2\alpha}^{xyy}$ and $F_{2\alpha}^{yxx}$, because $\partial_{b}d_{\alpha}$ vanishes at the band edge,  this indicates that
they also do not contribute to the shift current tensor when the optical frequency
exactly matches the band gap, i.e., $\omega=E_{g;\alpha}\equiv2\text{min}
\{d_{\alpha}\}=2\sqrt{\Delta_{\alpha;x}^{2}+\Delta_{\alpha;y}^{2}+m_{\alpha}^{2}}$.
Near the band edge, leading order terms of $F_{2\alpha}^{xyy}$ and $F_{2\alpha}^{yxx}$
are found to be linear in momentum, thus their contributions can be safely neglected in this regime.
For  $F_{1\alpha}^{abb}$, a straightforward calculation reveals $F_{1\alpha}^{abb}=F_{1\alpha}^{abb(0)}+\mathcal{O}(\bq)$,
where $F_{1\alpha}^{abb(0)}$ represents the zeroth-order term in momentum and
\begin{widetext}
\begin{eqnarray}
F_{1\alpha}^{xxx(0)}&=&(\Delta_{\alpha;x}\lambda_{\alpha;x}-m_{\alpha}v_{\alpha;x})B_{\alpha;xx}+
\Delta_{\alpha;y}(C_{\alpha;xx}v_{\alpha;x}-A_{\alpha;xx}\lambda_{x}),\nonumber\\
F_{1\alpha}^{yyy(0)}&=&(m_{\alpha}v_{\alpha;y}-\Delta_{\alpha;y}\lambda_{\alpha;y})A_{\alpha;yy}+\Delta_{\alpha;x}
(B_{\alpha;yy}\lambda_{\alpha;y}-C_{\alpha;yy}v_{\alpha;y}),\nonumber\\
F_{1\alpha}^{yxx(0)}&=&(\Delta_{\alpha;x}\lambda_{\alpha;x}-m_{\alpha}v_{\alpha;x})B_{\alpha;xy}+
\Delta_{\alpha;y}(C_{\alpha;xy}v_{\alpha;x}-A_{\alpha;xy}\lambda_{\alpha;x}), \nonumber\\
F_{1\alpha}^{xyy(0)}&=& (m_{\alpha}v_{\alpha;y}-\Delta_{\alpha;y}\lambda_{\alpha;y})A_{\alpha;xy}+\Delta_{\alpha;x}(B_{\alpha;xy}\lambda_{\alpha;y}-C_{\alpha;xy}v_{\alpha;y}).\label{leading}
\end{eqnarray}
\end{widetext}
As $\Delta_{\alpha;x}$, $\Delta_{\alpha;y}$ and $m_{\alpha}$ change their signs simultaneously,
it is readily seen that all components of $F_{1\alpha}^{abb(0)}$ will reverse their signs across the critical point.

Applying the formula in Eq.(\ref{formula}), we find that for optical frequency close
to the band gap ($\mu=0$ in
this work),
\begin{eqnarray}
\sigma^{abb}_{\alpha}(\omega)&\simeq&-\frac{e^{3}F_{1\alpha}^{abb(0)}}{2\hbar^{3}\omega^{2}\bar{v}_{\alpha}}
\tanh\frac{\hbar\omega}{4k_{B}T}\Theta(\hbar\omega-E_{g;\alpha}),\label{tensor}
\end{eqnarray}
where $\bar{v}_{\alpha}=\sqrt{v_{\alpha;x}^{2}v_{\alpha;y}^{2}
+\lambda_{\alpha;x}^{2}v_{\alpha;y}^{2}+v_{\alpha;x}^{2}\lambda_{\alpha;y}^{2}}$,  and $\Theta(x)$
is the Heviside step function.
For the convenience of discussion, we name $\sigma^{abb}_{\alpha}(\omega)$ with
$\omega=E_{g;\alpha}$ {\it band-edge shift current tensor}.
Two remarkable features of the band-edge shift current tensor can immediately be read from Eq.(\ref{tensor}):
(i) No matter what realistic value the temperature takes, all nonvanishing components will reverse their signs across the TPT;
(ii) All nonvanishing components have a discontinuous jump across the TPT,
with the discontinuous jump inversely proportional to the temperature, and going divergent in the zero-temperature limit.
Apparently, the singular sign-reversal behavior of band-edge shift current tensor
can be easily detected in experiments, thus it can be applied as a sensitive approach to determine the critical points of TPTs,
even for the class of TPTs driven by temperature\cite{Garate2013,Wiedmann2015,Antonius2016,Monserrat2016,Kadykov2018}.
Noteworthily, although here $H_{\alpha}(\bq)$ only describes
TPTs with Chern number jump $|\Delta C_{\alpha}|=1$,
the singular sign-reversal behavior also appears for more unusual ones
with $|\Delta C_{\alpha}|\geq2$ as nonvanishing band-edge shift current tensor must be
proportional to the zeroth order momentum terms in the low-energy Hamiltonian,
which, as we have analyzed above, will change their signs across the TPT. Therefore,
in the absence of inversion symmetry,
the singular sign-reversal behavior is expected to hold for various TPTs allowed by class A in 2D.

As $H_{\alpha}(\bq)$ may have symmetry-related partners, let us analyze the effects of the allowed symmetries to the shift
current tensor. In this work, for spatial symmetries, we confine ourselves to the symmorphic ones for simplicity and leave the more complicated nonsymmorphic symmetries for future study.
Clearly, in 2D the absence of inversion symmetry directly rules out
the $C_{2}$, $C_{4}$ and $C_{6}$ rotation symmetry, as well as the existence of
two mirror symmetries with respect to two orthogonal mirror planes. Therefore, 2D inversion asymmetric
insulators can at most simultaneously have  $C_{3}$ rotation symmetry,  TRS and certain mirror
symmetries whose mirror planes are not orthogonal.
Clearly, shift current tensors from the band edges related by $C_{3}$ rotation symmetry have
to be equal. Further analysis according to Eqs.(\ref{general}) and (\ref{formula})
reveals that shift current tensors from the band edges related by
TRS are equal, but are opposite for mirror symmetry.
Since the effect of TRS is a doubling of shift current tensors,
it becomes clear that the singular sign-reversal behavior of band-edge shift current tensor
also holds for TPTs allowed by class AII in 2D.

\begin{figure}
\subfigure{\includegraphics[width=7.5cm, height=2.5cm]{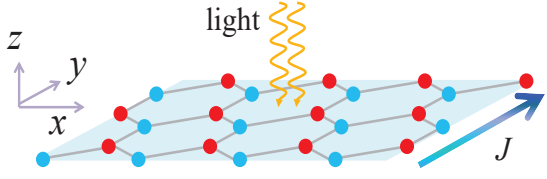}}
\caption{ A schematic of the honeycomb-lattice system. Without inversion symmetry, a photocurrent
will be generated when a beam of linearly polarized light is incident perpendicular to
the lattice plane.}  \label{sketch}
\end{figure}

{\it Concrete model.---} In the following we take the Kane-Mele model
as a concrete example to demonstrate the above general analysis. A schematic diagram
of the system is presented in Fig.\ref{sketch}.
The Hamiltonian reads\cite{kane2005a, kane2005b}
\begin{equation}
H=t\sum_{\langle ij\rangle}c_{i}^{\dag}c_{j}+i\lambda_{\rm so}\sum_{\langle\langle ij\rangle\rangle}\nu_{ij}c_{i}^{\dag}s^{z}c_{j}
+\lambda_{v}\sum_{i}\xi_{i}c_{i}^{\dag}c_{i},
\end{equation}
in which the three terms in sequence refer to the nearest-neighbour hopping,
spin-orbit coupling and stagger potential ($\xi_{i}=1 (-1)$ if $i\in A(B)$ sublattices)
on a honeycomb lattice, respectively.  The spin-orbit coupling
is related to the second-nearest-neighbour hopping, and when the hopping trajectory is anticlockwise (clockwise),
$v_{ij}=1 (-1)$. The presence of stagger potential breaks the crucial inversion symmetry.

In momentum space,
$H=\sum_{k}c_{k}^{\dag}H(\bk)c_{\bk}$ with $c_{\bk}=(c_{A,\bk\uparrow},c_{B,\bk\uparrow},
c_{A,\bk\downarrow},c_{B,\bk\downarrow})^{T}$ and
\begin{eqnarray}
H(\bk)&=&2\lambda_{\rm so}(2\sin\frac{\sqrt{3}k_{x}a}{2}\cos\frac{3k_{y}a}{2}-\sin\sqrt{3}k_{x}a)\tau_{z}s_{z}\nonumber\\
&&+\lambda_{v}\tau_{z}+t(\cos k_{y}a+2\cos\frac{\sqrt{3}k_{x}a}{2}\cos\frac{k_{y}a}{2})\tau_{x}\nonumber\\
&&+t(\sin k_{y}a-2\cos\frac{\sqrt{3}k_{x}a}{2}\sin\frac{k_{y}a}{2})\tau_{y},\label{HH}
\end{eqnarray}
where $\tau_{i}$ with $i=\{x,y,z\}$ are Pauli matrices acting on the sublattice space,
and $s_{z}$ is a Pauli matrix acting on the spin space.
For brevity of notation, below we set the lattice constant $a=1$ unless otherwise specified.
It is easy to check that a $2\pi/3$ rotation of the momentum, $(k_{x},k_{y})\rw(-k_{x}+\sqrt{3}k_{y}, -\sqrt{3}k_{x}-k_{y})/2$,
leaves the Hamiltonian intact, indicating the existence of
$C_{3}$ rotation symmetry. The Hamiltonian is also invariant under time-reversal operation
and mirror reflection about the $k_{x}=0$ plane,
i.e., $\mathcal{T}H(\bk)\mathcal{T}^{-1}=H(-\bk)$, $\mathcal{M}H(k_{x}, k_{y})\mathcal{M}^{-1}=H(-k_{x}, k_{y})$, with the time-reversal operator $\mathcal{T}=is_{y}\mathcal{K}$
($\mathcal{K}$ the complex conjugate operator) and the mirror reflection operator $\mathcal{M}=s_{x}$.

The Kane-Mele model belongs to class AII and is well-known to host two topologically distinct phases ($Z_{2}$ classification)\cite{kane2005a, kane2005b}.
For $|\lambda_{v}|<|\sqrt{3}\lambda_{\rm so}|$,  the model describes a quantum spin Hall insulator (or say topological insulator)
with helical gapless modes on the boundary, while for $|\lambda_{v}|>|\sqrt{3}\lambda_{\rm so}|$,
it describes a normal insulator without gapless boundary modes.   $|\lambda_{v}|=|\sqrt{3}\lambda_{\rm so}|$ is the critical point at which
the band gap is closed and TPT takes place.

As in this work the band-edge shift current tensor is of central interest,
below we also focus on the low-energy Hamiltonian around the band edge at first.
The band edges of this model
are located at the two valleys $\bK=(-4\pi/3\sqrt{3},0)$ and
$\bK'=(4\pi/3\sqrt{3},0)$. A short calculation reveals that the low-energy Hamiltonian is given by $H=\sum_{\chi,s;q}\oplus \Psi_{\chi,s;\bq}^{\dag}H_{\chi,s}(\bq)\Psi_{\chi,s;\bq}$,
where $\Psi_{\chi,s;\bq}=(c_{A,\bq s}, c_{B,\bq s})^{T}$ and $H_{\alpha,s}(\bq)=\bd_{\chi,s}(\bq)\cdot \tau$ with
\begin{eqnarray}
d_{\chi,s;x}(\bq)&=&\frac{3t}{2}(\chi q_{x}+\frac{1}{4}q_{x}^{2}-\frac{1}{4}q_{y}^{2}),\nonumber\\
d_{\chi,s;y}(\bq)&=&\frac{3t}{2}(q_{y}-\frac{1}{2}\chi q_{x}q_{y}),\nonumber\\
d_{\chi,s;z}(\bq)&=&m_{\chi,s}+\frac{9\sqrt{3}}{4}\lambda_{\rm so}\chi s(q_{x}^{2}+q_{y}^{2}),
\end{eqnarray}
where $\chi=1 (-1)$ for  $K$($K'$) valley, and $s=1 (-1)$ for up(down) spin.
$m_{\chi,s}=-3\sqrt{3}\lambda_{\rm so}\chi s+\lambda_{v}$ denotes
the Dirac mass.
Following Eqs.(\ref{formula}) and \ref{leading}, we find
$\sigma_{\chi,s}^{xxx}$ and $\sigma_{\chi,s}^{xyy}$ vanish identically for every choice of $(\chi,s)$,
thus $\sigma^{xxx}=\sum_{\chi,s}\sigma_{\chi,s}^{xxx}$ and $\sigma^{xyy}=\sum_{\chi,s}\sigma_{\chi,s}^{xyy}$
are both equal to zero, consistent with the fact that the full Hamiltonian is mirror symmetric about the $k_{x}=0$ plane.
For the remaining two components, we find that for optical frequency close
to the band gap\cite{shiftcurrent},
\begin{eqnarray}
\sigma_{\chi,s}^{yyy}(\omega)&\simeq&-\frac{e^{3}am_{\chi,s}}{4\hbar^{3}\eta_{\chi,s}\omega^{2}}
\tanh\frac{\hbar\omega}{4k_{B}T}\Theta(\hbar\omega-2|m_{\chi,s}|),\nonumber\\
\sigma_{\chi,s}^{yxx}(\omega)&\simeq&\frac{e^{3}am_{\chi,s}}{4\hbar^{3}\eta_{\chi,s}\omega^{2}}
\tanh\frac{\hbar\omega}{4k_{B}T}\Theta(\hbar\omega-2|m_{\chi,s}|),\label{expression}
\end{eqnarray}
where $\eta_{\chi,s}=1+2\sqrt{3}\lambda_{\rm so}m_{\chi,s}\chi s/t^{2}$ is a dimensionless
quantity. We have restored the lattice constant based on dimensional analysis. Once $\lambda_{v}\neq0$,
$\sigma^{yyy}=\sum_{\chi,s}\sigma_{\chi,s}^{yyy}$ and $\sigma^{yxx}=\sum_{\chi,s}\sigma_{\chi,s}^{yxx}$
will take a finite value for optical frequency above the band gap.
For the general case with finite $\lambda_{\rm so}$ and $\lambda_{v}$, the Dirac mass $m_{\chi,s}$
will have two different values, and only
the two contributions related to the smaller Dirac mass are relevant to band edges. For the convenience of discussion, below we consider
both  $\lambda_{\rm so}$ and $\lambda_{v}$ are positive, then the smaller Dirac mass takes the value $\lambda_{v}-3\sqrt{3}\lambda_{\rm so}$.
For frequency exactly matching the band gap, it is straightforward to find
\begin{eqnarray}
\sigma^{yyy}(\omega=2|m|)&=&-\frac{e^{3}a}{8\hbar\eta m}\tanh\frac{|m|}{2k_{B}T},   \nonumber\\
\sigma^{yxx}(\omega=2|m|)&=&\frac{e^{3}a}{8\hbar\eta m}\tanh\frac{|m|}{2k_{B}T},\label{result}
\end{eqnarray}
where $m=\lambda_{v}-3\sqrt{3}\lambda_{\rm so}$, and $\eta=1+2\sqrt{3}\lambda_{\rm so}m/t^{2}$.
Results clearly demonstrate that the band-edge shift current tensors, so too the band-edge shift current, will
reverse their signs in a singular way across the TPT.
It is noteworthy that if only one spin degree of freedom is considered, the Kane-Mele model reduces to the Haldane model (belongs to class A)
in which TRS is absent and
the TPTs are between a quantum anomalous Hall insulator (or say Chern insulator)
with Chern number $|C|=1$ and a normal insulator with $C=0$\cite{Haldane1988}.
Apparently,  the singular sign-reversal behavior of band-edge shift current tensor holds for the TPTs
of the Haldane model.

Before ending this section,  let us give further discussions on the shift current of this concrete model.
As only $\sigma^{yyy}$ and $\sigma^{yxx}$
take nonzero values when the optical frequency is above the band gap,
the shift current will be generated along the $y$ direction when a beam of linearly polarized light is incident
perpendicular to the system, as illustrated in Fig.\ref{sketch}.
Furthermore, as $\sigma^{yyy}\simeq-\sigma^{yxx}$,
the sign difference provides a knob to tune the strength and direction
of the shift current through the polarization of the light,
potentially allowing novel applications in optoelectronics.
Interestingly, when the polarization is bound in the $x$ direction,
the shift current is purely a nonlinear Hall current, i.e., the current flows in the direction perpendicular
to the optical field. As  is originated from interband processes,
this nonlinear Hall effect is distinct from the one induced by Berry curvature dipole\cite{Moore2010dipole,Sodemann2015dipole,Zhang2017dipole} which is
an intraband effect.

\begin{figure}
\subfigure{\includegraphics[width=7cm, height=5cm]{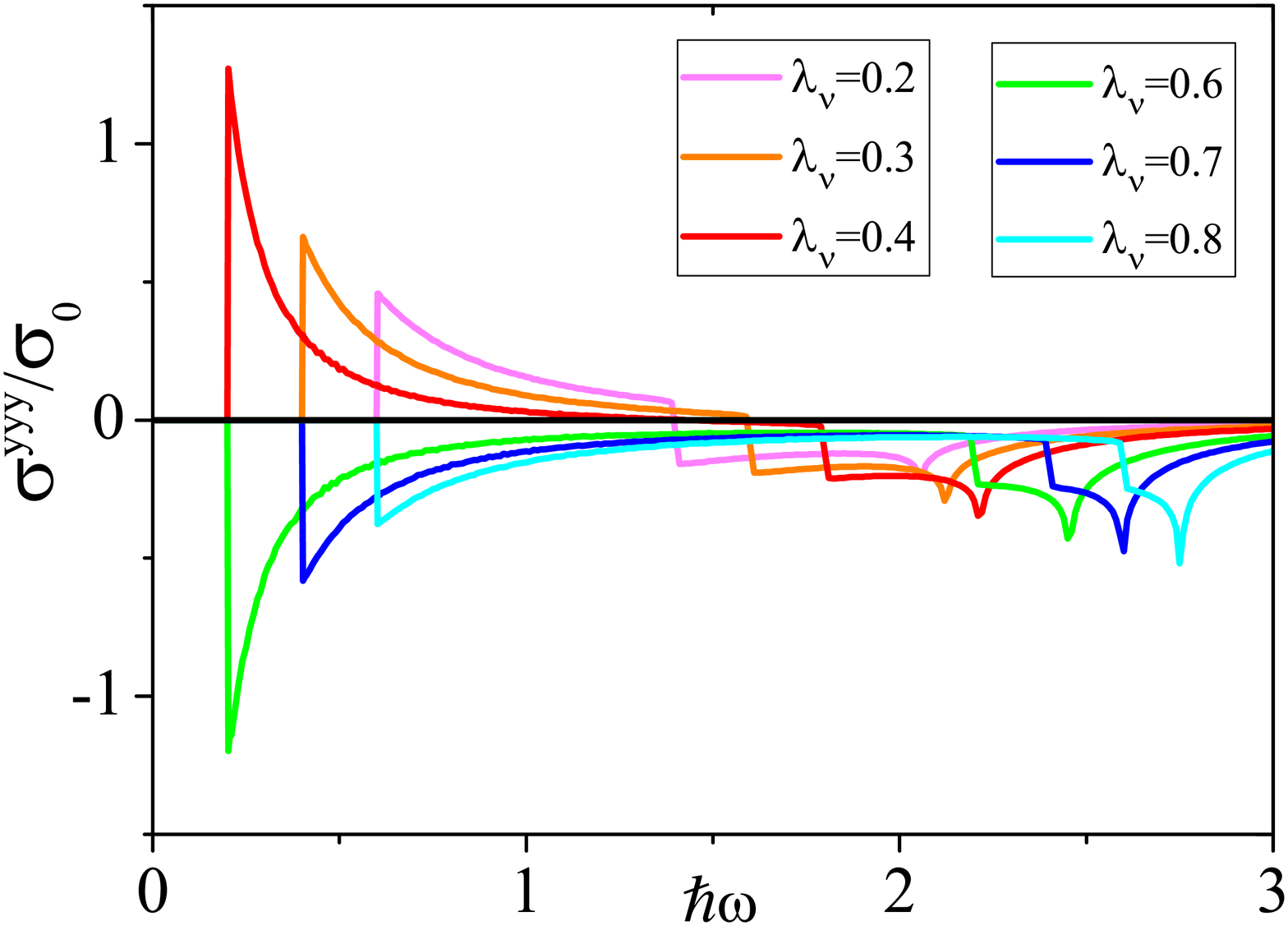}}
\caption{ The nonvanishing longitudinal component of shift current tensor for the Kane-Mele model. We set $t=1$ as
the energy unit, other parameters are: $\lambda_{\rm so}=\sqrt{3}/18$, $\sigma_{0}=(e^{3}a/\hbar)\tanh(\hbar\omega/4k_{B}T)$.
For the parameters chosen, $\lambda_{v}=0.5$ is a critical point of TPT.
The sign-reversal behavior of band-edge shift current tensor associated with TPT is clearly seen.
}  \label{puresc}
\end{figure}

For completeness, Fig.\ref{puresc} presents the shift current tensor
of the full Hamiltonian (\ref{HH}). As we found that $\sigma^{yyy}\simeq-\sigma^{yxx}$ holds
even for frequency much larger than the band gap, here only $\sigma^{yyy}$ is presented.  The result clearly
demonstrates the sign-reversal behavior of band-edge shift current tensor across the TPT.
Furthermore, we also
verified numerically that the introduction of Rashba spin-orbit coupling to the Kane-Mele model, which keeps the TRS but
breaks the spin conservation, does not change the sign-reversal behavior\cite{shiftcurrent},
indicating that this remarkable behavior holds no matter whether the spin conserves or not.

{\it Experimental considerations and conclusions.---} Though the relevance to many 2D inversion asymmetric  materials\cite{Yan2012review,Ando2013review,Ren20106review},
here we suggest two classes of materials to test our predictions. The first class of materials are  monolayer group-IV
elements, including silicene\cite{liu2011si,Liu2011si2}, germanene\cite{liu2011si,Liu2011si2}, stanene\cite{xu2013tin}, and their alloys, whose underlying topological properties are described by the very Kane-Mele
model. Owing to their buckled structures,
their band gaps can be continuously tuned by gate voltage\cite{Ni2012,Drummond2012,Ezawa2012} or strain\cite{Molle2017}, thus
continuous TPTs in this class of materials can be achieved. The second class of
materials are structural inversion asymmetric quantum wells. Noteworthily,
it has been experimentally confirmed that for sufficient thickness, the
AlSb/InAs/GaSb/AlSb quantum well is a small gap topological insulator\cite{Knez2011qsh}.
For this quantum well,  TPT  can also be continuously tuned by gate voltage\cite{Liu2008qsh},
thus our proposal can immediately be tested.

The shift current can be detected by
short-circuit current measurements\cite{Fridkin2001bpve,Nakamura2017,Osterhoudt2017}.
While tracking the evolution of band-edge shift current tensor across TPTs can irrefutably determine
the critical points, in experiments it is more practical to keep the frequency of optical field fixed
at a sufficiently small value, and detect the sign-reversal behavior of shift current only.
As an estimation,  we consider  the frequency $\hbar\omega=1$ meV, the field strength $\mathcal{E}=10^{5}$ V/m,
the temperature $T=10$K, the lattice constant $a=4${\AA}, and a change of Dirac mass from $m=0.01$ meV to
$-0.01$ meV,  then according to Eq.(\ref{expression}), it is readily found that the shift current will change from
$J\sim 0.1$ A/cm to $-0.1$ A/cm. Such a notable sign-reversal behavior can be easily detected in experiments.

In conclusion, we have demonstrated that for 2D inversion asymmetric insulators, with and without TRS,
a measurement of shift current can determine the critical points of
various TPTs precisely, even for temperature-driven ones\cite{Garate2013,Wiedmann2015,Antonius2016,Monserrat2016,Kadykov2018},
hopefully paving the way for future research on 2D TPTs.
This work may also stimulate studies of other nonlinear effects across TPTs, as well as
further exploration of the effects of interaction and disorder to such nonlinear effects.



{\it Acknowledgements.---}
The author would like to thank Shuai Yin and Wen Huang for
helpful suggestions on the manuscript.
The author would also like to acknowledge the support by a startup grant at Sun Yat-sen University,
and   express his thanks to Institute for Advanced Study, Tsinghua University,
where partial of this work was done.





\bibliography{dirac}

\widetext
\clearpage
\begin{center}
\textbf{\large Supplemental Material }\\
\vspace{4mm}
{ Zhongbo Yan}\\
\vspace{2mm}
{\em \small
School of Physics, Sun Yat-sen University, Guangzhou, 510275, China
}
\end{center}

\setcounter{equation}{0}
\setcounter{figure}{0}
\setcounter{table}{0}
\makeatletter
\renewcommand{\theequation}{S\arabic{equation}}
\renewcommand{\thefigure}{S\arabic{figure}}
\renewcommand{\bibnumfmt}[1]{[S#1]}

This supplemental material contains three parts: (I) The derivation of Eq.(\ref{formula}) in the main text;
(III) Some details of the calculation of shift current tensor, and a comparison between the
results from the full Hamiltonian and those from the low-energy
continuum Hamiltonian; (III) Demonstration
of the sign-reversal behavior of band-edge shift current tensor for spin nonconserving case.

\section{I. The derivation of Eq.(\ref{formula}) in the main text}

The shift current is a second-order optical effect
with the induced direct current proportional to the square of the optical field, i.e., $J_{a}=\sigma^{abc}(\omega)
\mathcal{E}_{b}(\omega)\mathcal{E}_{c}(-\omega)$. For linearly polarized light and in the independent particle approximation,
the shift current tensor $\sigma^{abc}$ is given by\cite{Sipe2000bpve}
\begin{eqnarray}
\sigma^{abb}(\omega)=\frac{2\pi e^{3}}{\hbar}\int\frac{d^{D}k}{(2\pi)^{D}}\sum_{\alpha\beta}
f_{\alpha\beta}|r_{\alpha\beta}^{b}|^{2}R_{\alpha\beta}^{a}
\delta(\hbar\omega-E_{\beta\alpha}),\quad\label{sgeneral}
\end{eqnarray}
where  $r_{\alpha\beta}^{a}=i\langle u_{\alpha}|\partial_{a}u_{\beta}\rangle$ ($\partial_{a}$ is a shorthand
notation of $\frac{\partial}{\partial k_{a}}$) with $\alpha\neq\beta$
is the interband Berry connection;
$f_{\alpha\beta}=f_{\alpha}-f_{\beta}$ with $f_{\alpha}=1/(1+\text{exp}[(E_{\alpha}(\bk)-\mu)/k_{\rm B}T])$ the Fermi-Dirac distribution
function, where $\mu$ is the chemical potential, $T$ is the temperature, and $k_{\rm B}$ is the Boltzmann constant; $E_{\beta\alpha}=E_{\beta}(\bk)-E_{\alpha}(\bk)$
represents the energy difference between two bands labeled by $\alpha$ and $\beta$ at momentum $\bk$;
$R_{\alpha\beta}^{a}=-\partial_{a}\text{arg}(r_{\alpha\beta}^{b})+\xi_{\alpha\alpha}^{a}-\xi_{\beta\beta}^{a}$ is
known as the shift vector which is related to the electric polarization between bands\cite{Fregoso2017shift}; $\xi_{\alpha\alpha}^{a}=i\langle u_{\alpha}|\partial_{a}u_{\alpha}\rangle$
is the intraband Berry connection. Although the shift vector involves the
gauge-dependent Berry connection, the formula in Eq.(\ref{sgeneral}) is
gauge-invariant.

Let us focus on the two-band case. For a two-band Hamiltonian, it can always be expressed  in terms of the Pauli matrices as
\begin{eqnarray}
H(\bk)=\bd(\bk)\cdot\tau+\epsilon(\bk)\mathbf{I},
\end{eqnarray}
where $\bd(\bk)=(d_{x}(\bk),d_{y}(\bk),d_{z}(\bk))$, $\tau=(\tau_{x},\tau_{y},\tau_{z})$ are Pauli
matrices, and $\mathbf{I}$ is
the rank-2 unit matrix. Correspondingly, the energy spectra read
\begin{eqnarray}
E_{\pm}(\bk)=\epsilon(\bk)\pm d(\bk),
\end{eqnarray}
and the eigenvectors take the form
\begin{eqnarray}
|u_{+}(\bk)\rangle=\left(
                     \begin{array}{c}
                       \cos\frac{\theta_{\bk}}{2} \\
                       \sin\frac{\theta_{\bk}}{2}e^{i\phi_{\bk}} \\
                     \end{array}
                   \right),\quad
|u_{-}(\bk)\rangle=\left(
                     \begin{array}{c}
                       \sin\frac{\theta_{\bk}}{2}e^{-i\phi_{\bk}} \\
                       -\cos\frac{\theta_{\bk}}{2} \\
                     \end{array}
                   \right),
\end{eqnarray}
where $d(\bk)=\sqrt{\sum_{i=x,y,z}d_{i}(\bk)^{2}}$, $\theta_{\bk}=\arctan\sqrt{d_{x}^{2}(\bk)+d_{y}^{2}(\bk)}/d_{z}(\bk)$, and
$\phi_{\bk}=\arctan d_{y}(\bk)/d_{x}(\bk)$ (for brevity of notation,
in this section we no longer write down the $\bk$-dependence explicitly).
According to the eigenvectors, the interband and intraband Berry connections are
given by
\begin{eqnarray}
r_{-+}^{b}&=&i\langle u_{-}|\partial_{b}u_{+}\rangle\nonumber\\
&=&-\frac{ie^{i\phi_{k}}}{2}(\partial_{b}\theta_{\bk}+i\sin\theta_{\bk}\partial_{b}\phi_{\bk})=(r_{+-}^{b})^{*},\nonumber\\
\xi_{--}^{a}&=&i\langle u_{-}|\partial_{a}u_{-}\rangle
=\sin^{2}\frac{\theta_{\bk}}{2}\partial_{a}\phi_{\bk},\nonumber\\
\xi_{++}^{a}&=&i\langle u_{+}|\partial_{a}u_{+}\rangle
=-\sin^{2}\frac{\theta_{\bk}}{2}\partial_{a}\phi_{\bk}.
\end{eqnarray}
Thus, $|r_{-+}^{b}|^{2}=[(\partial_{b}\theta_{\bk})^{2}+\sin^{2}\theta_{\bk}(\partial_{b}\phi_{\bk})^{2}]/4$,
and the shift vector is given by
\begin{eqnarray}
R_{-+}^{a}&=&-\partial_{a}\text{arg}(r_{-+}^{b})+\xi_{--}^{a}-\xi_{++}^{a}\nonumber\\
&=&-\partial_{a}\phi_{\bk}-\partial_{a}\arctan \frac{\sin\theta_{\bk}\partial_{b}\phi_{\bk}}{\partial_{b}\theta_{\bk}}+2\sin^{2}\frac{\theta_{\bk}}{2}\partial_{a}\phi_{\bk}\nonumber\\
&=&-\cos\theta_{\bk}\partial_{a}\phi_{\bk}-\partial_{a}\arctan \frac{\sin\theta_{\bk}\partial_{b}\phi_{\bk}}{\partial_{b}\theta_{\bk}}.
\end{eqnarray}
By using the expression of $\theta_{k}$ and $\phi_{\bk}$,
a straightforward calculation reveals
\begin{eqnarray}
|r_{-+}^{b}|^{2}=\frac{1}{8d^{4}}\sum_{ij}(d_{i}\partial_{b}d_{j}-d_{j}\partial_{b}d_{i})^{2}.
\end{eqnarray}
It is readily seen that $|r_{-+}^{b}|^{2}$ is invariant under the cyclic changes, $\{d_{x}\rw d_{y},d_{y}\rw d_{z},
d_{z}\rw d_{x}\}$ and $\{d_{x}\rw d_{z},d_{y}\rw d_{x},
d_{z}\rw d_{y}\}$. Meanwhile, the shift current tensor is a physical quantity, such a cyclic change will also not affect
its result.
By using this cyclic property, it is straightforward to find
\begin{eqnarray}
\int\cos\theta_{\bk}\partial_{a}\phi_{\bk}&=&\int\frac{d_{z}(d_{x}\partial_{a}d_{y}-d_{y}\partial_{a}d_{x})}{d(d_{x}^{2}+d_{y}^{2})}\nonumber\\
&=&\int\frac{d_{z}(d_{x}\partial_{a}d_{y}-d_{y}\partial_{a}d_{x})+\text{cyclic changes}}{d(d_{x}^{2}+d_{y}^{2})+\text{cyclic changes} }\nonumber\\
&=&\int\frac{\bd\cdot(\bd\times\partial_{a}\bd)}{2d^{3} }=0.\label{cyclic}
\end{eqnarray}
Thus,
the shift current tensor is given by
\begin{eqnarray}
\sigma^{abb}&=&\frac{2\pi e^{3}}{\hbar}\int\frac{d^{D}k}{(2\pi)^{D}}
f_{-+}|r_{-+}^{b}|^{2}R_{-+}^{a}
\delta(\hbar\omega-2d)\nonumber\\
&=&-\frac{2\pi e^{3}}{\hbar}\int\frac{d^{D}k}{(2\pi)^{D}}
f_{-+}|r_{-+}^{b}|^{2}(\partial_{a}\arctan \frac{\sin\theta_{\bk}\partial_{b}\phi_{\bk}}{\partial_{b}\theta_{\bk}}+\text{cyclic changes})
\delta(\hbar\omega-2d)\nonumber\\
&=&\frac{\pi e^{3}}{2\hbar}\int\frac{d^{D}k}{(2\pi)^{D}}
\{[\sin\theta_{\bk}\partial_{b}\phi_{\bk}\partial_{ab}\theta_{\bk}-
\partial_{a}(\sin\theta_{\bk}\partial_{b}\phi_{\bk})\partial_{b}\theta_{\bk}
]+\text{cyclic changes}\}
f_{-+}\delta(\hbar\omega-2d).\nonumber
\end{eqnarray}
It is noteworthy that here the ``cyclic change'' means that the nominator and denominator of
$[\sin\theta_{\bk}\partial_{b}\phi_{\bk}\partial_{ab}\theta_{\bk}-
\partial_{a}(\sin\theta_{\bk}\partial_{b}\phi_{\bk})\partial_{b}\theta_{\bk}
]$ do a simultaneous cyclic change, like that in Eq.(\ref{cyclic}).
The integrand in the bracket can be rewritten in terms of the three components
of the $\bd$-vector. According to the expressions of $\theta_{\bk}$ and $\phi_{\bk}$, it is
readily found
\begin{eqnarray}
&&\sin\theta_{\bk}\partial_{b}\phi_{\bk}=\frac{d_{x}\partial_{a}d_{y}-d_{y}\partial_{a}d_{x}}{d\sqrt{d_{x}^{2}+d_{y}^{2}}},\nonumber\\
&&\partial_{b}\theta_{\bk}=\frac{(d_{x}\partial_{b}d_{x}+d_{y}\partial_{b}d_{y})d_{z}-(d_{x}^{2}+d_{y}^{2})\partial_{b}d_{z}}
{d^{2}\sqrt{d_{x}^{2}+d_{y}^{2}}},
\end{eqnarray}
then
\begin{eqnarray}
&&\sin\theta_{\bk}\partial_{b}\phi_{\bk}\partial_{ab}\theta_{\bk}-
\partial_{a}(\sin\theta_{\bk}\partial_{b}\phi_{\bk})\partial_{b}\theta_{\bk}\nonumber\\
&=&\frac{(d_{x}\partial_{b}d_{y}-d_{y}\partial_{b}d_{x})}
{d^{5}(d_{x}^{2}+d_{y}^{2})^{2}}\left\{d^{2}(d_{x}^{2}+d_{y}^{2})\left[(\partial_{a}d_{x}\partial_{b}d_{x}+
d_{x}\partial_{ab}d_{x}+\partial_{a}d_{y}\partial_{b}d_{y}+
d_{y}\partial_{ab}d_{y})d_{z}+(d_{x}\partial_{b}d_{x}+d_{y}\partial_{b}d_{y})\partial_{a}d_{z}\right.\right.\nonumber\\
&&\left.-2(d_{x}\partial_{a}d_{x}+d_{y}\partial_{a}d_{y})
\partial_{b}d_{z}-(d_{x}^{2}+d_{y}^{2})\partial_{ab}d_{z}\right]-2(d_{x}^{2}+d_{y}^{2})\left[(d_{x}\partial_{b}d_{x}+d_{y}\partial_{b}d_{y})d_{z}
-(d_{x}^{2}+d_{y}^{2})\partial_{b}d_{z}\right]\nonumber\\
&&\left.\times(d_{x}\partial_{a}d_{x}+d_{y}\partial_{a}d_{y}+d_{z}\partial_{a}d_{z})-d^{2}\left[(d_{x}\partial_{b}d_{x}+d_{y}\partial_{b}d_{y})d_{z}
-(d_{x}^{2}+d_{y}^{2})\partial_{b}d_{z}\right](d_{x}\partial_{a}d_{x}+d_{y}\partial_{a}d_{y})\right\}\nonumber\\
&&-\frac{\left[(d_{x}\partial_{b}d_{x}+d_{y}\partial_{b}d_{y})d_{z}
-(d_{x}^{2}+d_{y}^{2})\partial_{b}d_{z}\right]}
{d^{5}(d_{x}^{2}+d_{y}^{2})^{2}}\left\{d^{2}(d_{x}^{2}+d_{y}^{2})(\partial_{a}d_{x}\partial_{b}d_{y}+d_{x}\partial_{ab}d_{y}
-\partial_{a}d_{y}\partial_{b}d_{x}-d_{y}\partial_{ab}d_{x})\right.\nonumber\\
&&\left.-(d_{x}^{2}+d_{y}^{2})(d_{x}\partial_{b}d_{y}-d_{y}\partial_{b}d_{x})(d_{x}\partial_{a}d_{x}+d_{y}\partial_{a}d_{y}+d_{z}\partial_{a}d_{z})
-d^{2}(d_{x}\partial_{b}d_{y}-d_{y}\partial_{b}d_{x})(d_{x}\partial_{a}d_{x}+d_{y}\partial_{a}d_{y})\right\}.
\end{eqnarray}
We first consider the terms with second derivative, which
give
\begin{eqnarray}
P_{1}&=&\frac{1}
{d^{5}(d_{x}^{2}+d_{y}^{2})^{2}}\left\{d^{2}(d_{x}^{2}+d_{y}^{2})(d_{x}\partial_{b}d_{y}-d_{y}\partial_{b}d_{x})
\left[(d_{x}\partial_{ab}d_{x}+d_{y}\partial_{ab}d_{y})d_{z}-(d_{x}^{2}+d_{y}^{2})\partial_{ab}d_{z}\right]\right.\nonumber\\
&&\left.-d^{2}(d_{x}^{2}+d_{y}^{2})(d_{x}\partial_{ab}d_{y}-d_{y}\partial_{ab}d_{x})\left[(d_{x}\partial_{b}d_{x}+d_{y}\partial_{b}d_{y})d_{z}
-(d_{x}^{2}+d_{y}^{2})\partial_{b}d_{z}\right]\right\}\nonumber\\
&=&\frac{1}
{d^{3}(d_{x}^{2}+d_{y}^{2})}\left\{(d_{x}^{2}+d_{y}^{2})\left[(d_{x}\partial_{ab}d_{y}-d_{y}\partial_{ab}d_{x})\partial_{b}d_{z}
-(d_{x}\partial_{b}d_{y}-d_{y}\partial_{b}d_{x})\partial_{ab}d_{z}\right]\right.\nonumber\\
&&\left.+\left[(d_{x}\partial_{b}d_{y}-d_{y}\partial_{b}d_{x})(d_{x}\partial_{ab}d_{x}+d_{y}\partial_{ab}d_{y})
-(d_{x}\partial_{ab}d_{y}-d_{y}\partial_{ab}d_{x})(d_{x}\partial_{b}d_{x}+d_{y}\partial_{b}d_{y})\right]d_{z}\right\}\nonumber\\
&=&\frac{1}
{d^{3}}\left[(d_{x}\partial_{ab}d_{y}-d_{y}\partial_{ab}d_{x})\partial_{b}d_{z}
-(d_{x}\partial_{b}d_{y}-d_{y}\partial_{b}d_{x})\partial_{ab}d_{z}+
(\partial_{ab}d_{x}\partial_{b}d_{y}-\partial_{ab}d_{y}\partial_{b}d_{x})d_{z}\right]\nonumber\\
&=&\frac{1}
{d^{3}}\left[d_{x}(\partial_{ab}d_{y}\partial_{b}d_{z}-\partial_{ab}d_{z}\partial_{b}d_{y})
+d_{y}(\partial_{ab}d_{z}\partial_{b}d_{x}-\partial_{ab}d_{x}\partial_{b}d_{z})+
d_{z}(\partial_{ab}d_{x}\partial_{b}d_{y}-\partial_{ab}d_{y}\partial_{b}d_{x})\right]\nonumber\\
&=&\frac{\bd\cdot(\partial_{ab}\bd\times\partial_{b}\bd)}{d^{3}}.
\end{eqnarray}
It is apparent that such a form is invariant under the cyclic change.
Thus, the contribution of this part to the shift current tensor is
\begin{eqnarray}
&&\frac{\pi e^{3}}{2\hbar}\int\frac{d^{D}k}{(2\pi)^{D}}\frac{\bd\cdot(\partial_{ab}\bd\times\partial_{b}\bd)}{d^{3}}
f_{-+}\delta(\hbar\omega-2d)\nonumber\\
&&=\frac{\pi e^{3}}{2\hbar}\int\frac{d^{D}k}{(2\pi)^{D}}\frac{\bd\cdot(\partial_{ab}\bd\times\partial_{b}\bd)}{(\hbar\omega/2)^{3}}
f_{-+}\delta(\hbar\omega-2d)\nonumber\\
&&=\frac{4\pi e^{3}}{\hbar^{4}\omega^{3}}\int\frac{d^{D}k}{(2\pi)^{D}}\bd\cdot(\partial_{ab}\bd\times\partial_{b}\bd)
f_{-+}\delta(\hbar\omega-2d)\nonumber\\
&&=\sigma_{{\rm I}}^{abb}.
\end{eqnarray}
For the remaining parts, by using the cyclic change of $\bd$-vector and doing
some lengthy but straightforward calculations, we find the result is
\begin{eqnarray}
P_{2}&=&\frac{1}{d^{5}}\left[\sum_{ijk}\epsilon^{ijk}d_{j}^{2}\partial_{a}d_{i}\partial_{b}d_{j}\partial_{b}d_{k}
+\sum_{ijk}\epsilon^{ijk}d_{i}d_{j}\partial_{b}d_{i}(\partial_{a}d_{i}\partial_{b}d_{k}-\partial_{b}d_{i}\partial_{a}d_{k})\right]\nonumber\\
&=&\frac{1}{d^{5}}\left[\sum_{ijk}\epsilon^{ijk}d_{j}\partial_{a}d_{i}\partial_{b}d_{k}(d_{i}\partial_{b}d_{i}+d_{j}\partial_{b}d_{j})
-\sum_{ijk}\epsilon^{ijk}d_{j}\partial_{a}d_{k}(d_{i}\partial_{b}d_{i})\right]\nonumber\\
&=&\frac{1}{d^{5}}\left[\sum_{ijk}\epsilon^{ijk}d_{j}\partial_{a}d_{i}\partial_{b}d_{k}(d_{i}\partial_{b}d_{i}+d_{j}\partial_{b}d_{j})
-\sum_{ijk}\epsilon^{kji}d_{j}\partial_{b}d_{k}\partial_{a}d_{i}(d_{k}\partial_{b}d_{k})\right]\nonumber\\
&=&\frac{1}{d^{5}}\left[\sum_{ijk}\epsilon^{ijk}d_{j}\partial_{a}d_{i}\partial_{b}d_{k}(d_{i}\partial_{b}d_{i}
+d_{j}\partial_{b}d_{j}+d_{k}\partial_{b}d_{k})\right]\nonumber\\
&=&\frac{1}{d^{5}}\left[\sum_{ijk}\epsilon^{ijk}d_{j}\partial_{a}d_{i}\partial_{b}d_{k}(d\partial_{b}d)\right]\nonumber\\
&=&\frac{1}{d^{4}}\left[\sum_{ijk}\epsilon^{ijk}d_{j}\partial_{a}d_{i}\partial_{b}d_{k}(\partial_{b}d)\right]\nonumber\\
&=&-\frac{1}{d^{4}}\left[\bd\cdot(\partial_{a}\bd\times\partial_{b}\bd)(\partial_{b}d)\right]\nonumber\\
&=&\frac{2}{d}\Omega_{ab}(\partial_{b}d),
\end{eqnarray}
where $\Omega_{ab}=-\bd\cdot(\partial_{a}\bd\times\partial_{b}\bd)/2d^{3}$ is the Berry curvature
of the valence band. The contribution of this part to the shift current tensor is
\begin{eqnarray}
&&\frac{\pi e^{3}}{2\hbar}\int\frac{d^{D}k}{(2\pi)^{D}}\frac{2\Omega_{ab}(\partial_{b}d)}{d}
f_{-+}\delta(\hbar\omega-2d)\nonumber\\
&&=\frac{\pi e^{3}}{2\hbar}\int\frac{d^{D}k}{(2\pi)^{D}}\frac{2\Omega_{ab}(\partial_{b}d)}{\hbar\omega/2}
f_{-+}\delta(\hbar\omega-2d)\nonumber\\
&&=\frac{2\pi e^{3}}{\hbar^{2}\omega}\int\frac{d^{D}k}{(2\pi)^{D}}\Omega_{ab}(\partial_{b}d)
f_{-+}\delta(\hbar\omega-2d)\nonumber\\\nonumber\\
&&=\sigma_{{\rm II}}^{abb}.
\end{eqnarray}

The summation of $\sigma_{{\rm I}}^{abb}$ and $\sigma_{{\rm II}}^{abb}$
gives the formula in Eq.(\ref{formula}) of the main text.

\section{II. Shift current tensor of the low-energy continuum Hamiltonian}

In momentum space, the Kane-Mele model is given by $H=\sum_{k}c_{k}^{\dag}H(\bk)c_{\bk}$ with $c_{\bk}=(c_{A,\bk\uparrow},c_{B,\bk\uparrow},
c_{A,\bk\downarrow},c_{B,\bk\downarrow})^{T}$ and
\begin{eqnarray}
H(\bk)&=&2\lambda_{\rm so}(2\sin\frac{\sqrt{3}k_{x}a}{2}\cos\frac{3k_{y}a}{2}-\sin\sqrt{3}k_{x}a)\tau_{z}s_{z}\nonumber\\
&&+\lambda_{v}\tau_{z}+t(\cos k_{y}a+2\cos\frac{\sqrt{3}k_{x}a}{2}\cos\frac{k_{y}a}{2})\tau_{x}\nonumber\\
&&+t(\sin k_{y}a-2\cos\frac{\sqrt{3}k_{x}a}{2}\sin\frac{k_{y}a}{2})\tau_{y},\label{SHH}
\end{eqnarray}
where $\tau_{i}$ with $i=\{x,y,z\}$ are Pauli matrices acting on the sublattice space,
and $s_{z}$ is a Pauli matrix acting on the spin space. For brevity of notation,
below we set the lattice constant $a=1$ unless otherwise specified.
For this model,  the band edges are located at the two points $\bK=(-\frac{4\pi}{3\sqrt{3}},0)$ and
$\bK'=(\frac{4\pi}{3\sqrt{3}},0)$. By expanding the full Hamiltonian in Eq.(\ref{SHH}) around these two points  to second order
in momentum, the low-energy continuum Hamiltonian is given by $H(\bq)=\sum_{\chi,s}\oplus \bd_{\chi,s}(\bq)\cdot \mathbf{\tau}$ with
\begin{eqnarray}
d_{\chi,s;x}(\bq)&=&\frac{3t}{2}(\chi q_{x}+\frac{1}{4}q_{x}^{2}-\frac{1}{4}q_{y}^{2}),\nonumber\\
d_{\chi,s;y}(\bq)&=&\frac{3t}{2}(q_{y}-\frac{1}{2}\chi q_{x}q_{y}),\nonumber\\
d_{\chi,s;z}(\bq)&=&-3\sqrt{3}\lambda_{\rm so}\chi s[1-\frac{3}{4}(q_{x}^{2}+q_{y}^{2})]+\lambda_{v},
\end{eqnarray}
where $\chi=+1$ for $K$ valley, $\chi=-1$ for  $K'$ valley, $s=+1$ for up spin and $s=-1$ for down spin.
For the energy spectra, we also keep the momentum to second order,
\begin{eqnarray}
E_{\chi,s;\pm}(\bq)=\pm d_{\chi,s}(\bq)=\pm\sqrt{\frac{9\eta_{\chi,s}t^{2}}{4}(q_{x}^{2}+q_{y}^{2})+m_{\chi,s}^{2}},
\end{eqnarray}
where $m_{\chi,s}=-3\sqrt{3}\lambda_{\rm so}\chi s+\lambda_{v}$, and
$\eta_{\chi,s}=1+2\sqrt{3}\lambda_{\rm so}m_{\chi,s}\chi s/t^{2}$. For this low-energy
continuum Hamiltonian, we will consider $t\gg\lambda_{\rm so},  \lambda_{v}$
so that $\eta_{\chi,s}$ is always positive definite and the band edges are indeed located at
the two points $\bK$ and $\bK'$.

A short calculation reveals
\begin{eqnarray}
F_{1;\chi,s}^{yyy}&=&\bd_{\chi,s}(\bq)\cdot(\partial_{yy}\bd_{\chi,s}(\bq)\times\partial_{y}\bd_{\chi,s}(\bq))\nonumber\\
&=&\frac{9t^{2}}{8}(3\sqrt{3}\lambda_{\rm so}\chi s-\lambda_{v})+\mathcal{O}(\bq).
\end{eqnarray}
Near the band edge, all higher order terms contained in $\mathcal{O}(\bq)$ can
be safely neglected as $\bq\rw0$. For frequency close to the band gap,
\begin{eqnarray}
\sigma_{\chi,s}^{yyy}(\omega)&\simeq&\frac{4\pi e^{3}}{\hbar^{4}\omega^{3}}\int\frac{d^{2}q}{(2\pi)^{2}}\frac{9t^{2}}{8}(3\sqrt{3}\lambda_{\rm so}\chi s
-\lambda_{v})\delta(\hbar\omega-2\sqrt{\frac{9\eta_{\chi,s}t^{2}}{4}(q_{x}^{2}+q_{y}^{2})+m_{\chi,s}^{2}})\nonumber\\
&=&\frac{ e^{3}}{2\hbar^{4}\eta_{\chi,s}\omega^{3}}\int_{0}^{\infty}dx(3\sqrt{3}\lambda_{\rm so}\chi s
-\lambda_{v})
\delta(\hbar\omega-2\sqrt{x+m_{\chi,s}^{2}})\nonumber\\
&=&\frac{e^{3}}{4\hbar^{3}\eta_{\chi,s}\omega^{2}}(3\sqrt{3}\lambda_{\rm so}\chi s
-\lambda_{v})\Theta(\hbar\omega-2|m_{\chi,s}|)\nonumber\\
&=&-\frac{e^{3}m_{\chi,s}}{4\hbar^{3}\eta_{\chi,s}\omega^{2}}\Theta(\hbar\omega-2|m_{\chi,s}|).
\end{eqnarray}

For the off-diagonal component, we find
\begin{eqnarray}
F_{1;\chi,s}^{yxx}&=&\bd_{\chi,s}(\bq)\cdot(\partial_{yx}\bd_{\chi,s}(\bq)\times\partial_{x}\bd_{\chi,s}(\bq))\nonumber\\
&=&-\frac{9t^{2}}{8}(3\sqrt{3}\lambda_{\rm so}\chi s-\lambda_{v})+\mathcal{O}(\bq),\nonumber\\
F_{2;\chi,s}^{yxx}&=&(\hbar\omega)^{2}\Omega_{\chi,s;yx}\partial_{x}d_{\chi,s}/2=\mathcal{O}(\bq).\nonumber
\end{eqnarray}
Again, the terms contained in $\mathcal{O}(\bq)$ can
be safely neglected near the band edge. Thus for frequency close to the band gap,
\begin{eqnarray}
\sigma_{\chi,s}^{yxx}(\omega)&\simeq&\frac{4\pi e^{3}}{\hbar^{4}\omega^{3}}\int\frac{d^{2}q}{(2\pi)^{2}}[-\frac{9t^{2}}{8}(3\sqrt{3}\lambda_{\rm so}\chi s
-\lambda_{v})]\delta(\hbar\omega-2\sqrt{\frac{9\eta_{\chi,s}t^{2}}{4}(q_{x}^{2}+q_{y}^{2})+m_{\chi,s}^{2}})\nonumber\\
&=&-\frac{e^{3}}{4\hbar^{3}\eta_{\chi,s}\omega^{2}}(3\sqrt{3}\lambda_{\rm so}\chi s
-\lambda_{v})\Theta(\hbar\omega-2|m_{\chi,s}|)\nonumber\\
&=&\frac{e^{3}m_{\chi,s}}{4\hbar^{3}\eta_{\chi,s}\omega^{2}}\Theta(\hbar\omega-2|m_{\chi,s}|)\nonumber\\
&\simeq&-\sigma_{\chi,s}^{yyy}(\omega).
\end{eqnarray}

If we restore the lattice constant, simple dimensional analysis reveals
\begin{eqnarray}
\sigma_{\chi,s}^{yxx}(\omega)&\simeq&-\sigma_{\chi,s}^{yyy}(\omega)\nonumber\\
&\simeq&-\frac{e^{3}a}{4\hbar^{3}\eta_{\chi,s}\omega^{2}}(3\sqrt{3}\lambda_{\rm so}\chi s
-\lambda_{v})\Theta(\hbar\omega-2|m_{\chi,s}|).\nonumber
\end{eqnarray}

\begin{figure}
\subfigure{\includegraphics[width=7cm, height=5cm]{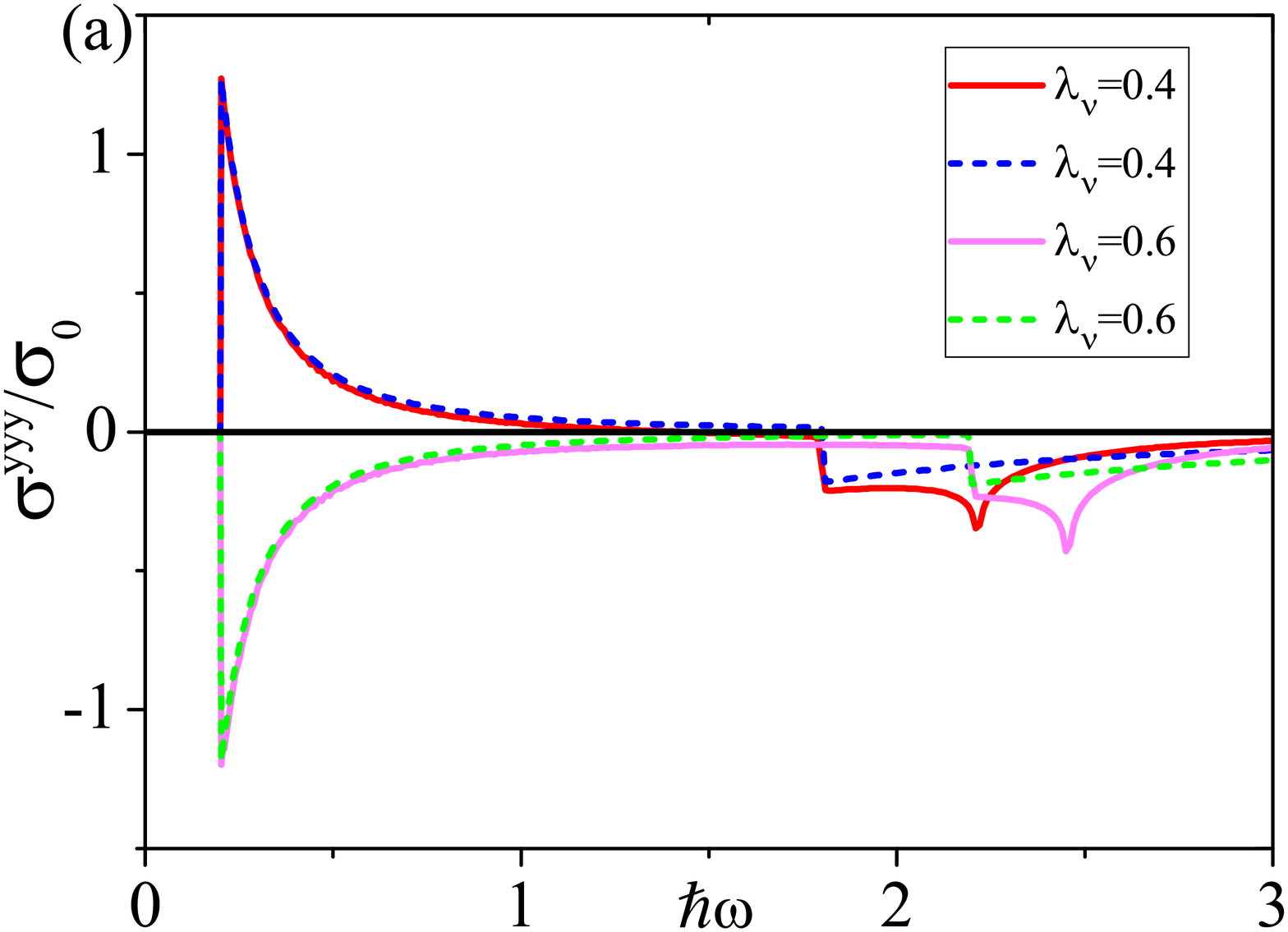}}
\subfigure{\includegraphics[width=7cm, height=5cm]{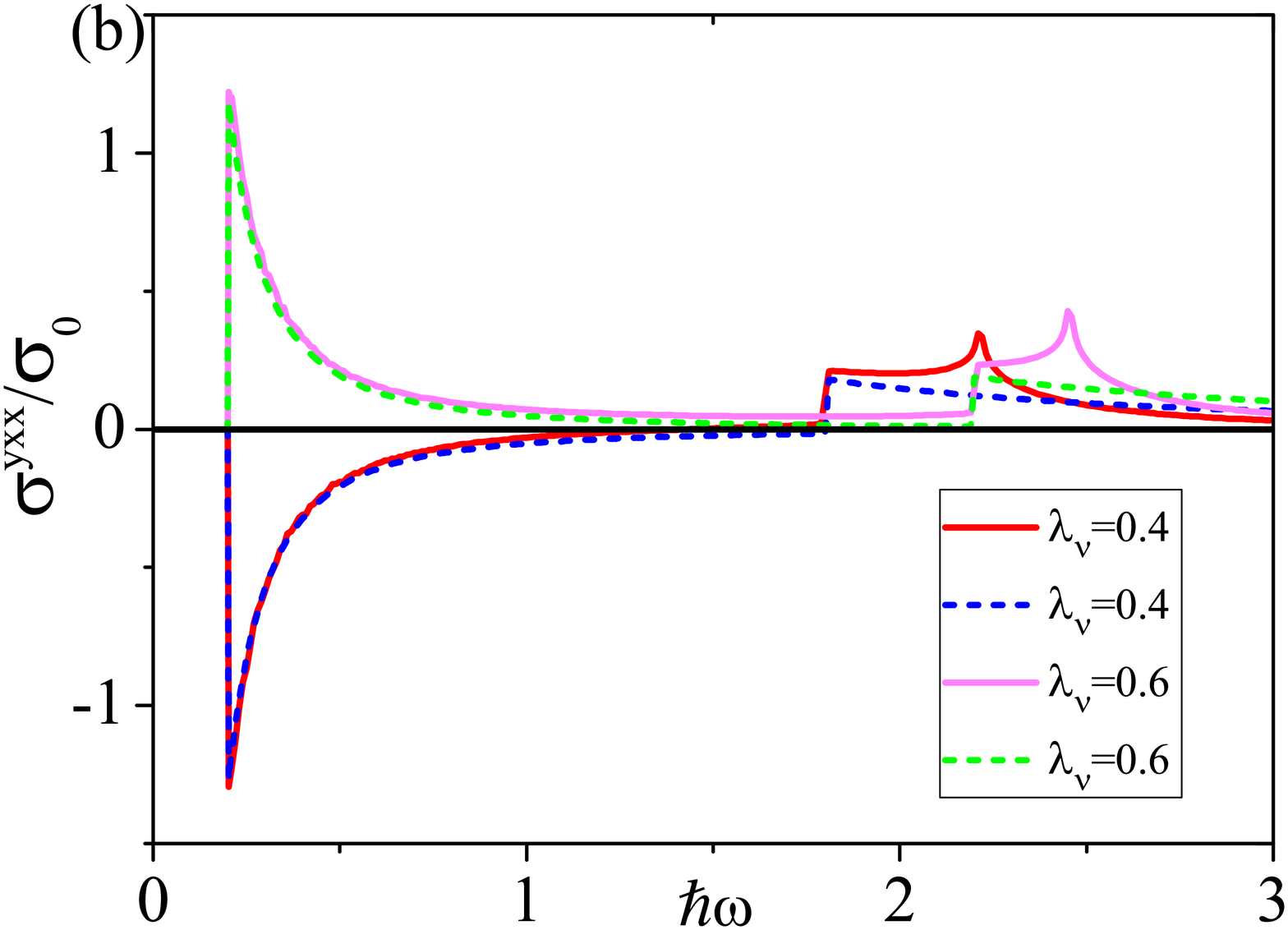}}
\caption{ The solid lines refer to shift current tensors obtained
from the full lattice Hamiltonian, while the dashed lines refer to
shift current tensors obtained from the low-energy continuum Hamiltonian. Parameters
$t=1$,  $\lambda_{\rm so}=\sqrt{3}/18$, (a) $\lambda_{v}=0.4$, and (b) $\lambda_{v}=0.6$.
The chemical potential $\mu$ and temperature $T$ are fixed to zero, correspondingly, $\sigma_{0}=e^{3}a/\hbar$.
For the parameters chosen, there is a critical point  at $\lambda_{v}=0.5$. }  \label{comparison}
\end{figure}

We present the results from the full Hamiltonian and those from the low-energy continuum Hamiltonian in Fig.\ref{comparison}
for a comparison.
In the figure, the solid lines refer to shift current tensors of the full
Hamiltonian, and the dashed lines refer to shift current tensors of
the low-energy continuum Hamiltonian under the same parameter condition,
it is remarkable that although only leading order terms are kept for the shift current tensors of the
low-energy continuum Hamiltonian, the results from the full Hamiltonian and those
from the low-energy continuum Hamiltonian  agree with each other very well in a
considerably broad range.

\section{III. Spin non-conserving case}

In this section, we demonstrate that the sign-reversal behavior of band-edge shift current tensor holds even when the spin in the Kane-Mele
model no longer conserves. For concreteness, we introduce the Rashba spin-orbit coupling to break the spin conversation,
which is given by
\begin{eqnarray}
H_{\rm R}=i\lambda_{\rm R}\sum_{\langle i,j\rangle}c^{\dag}_{i}(\mathbf{s}\times \hat{\bd}_{ij})_{z}c_{j},
\end{eqnarray}
where $\hat{\bd}_{ij}$ denotes the unit vector connecting $i$ and $j$. Then the full Hamiltonian in momentum space becomes
\begin{eqnarray}
H(\bk)&=&2\lambda_{\rm so}(2\sin\frac{\sqrt{3}k_{x}a}{2}\cos\frac{3k_{y}a}{2}-\sin\sqrt{3}k_{x}a)\tau_{z}s_{z}
+\lambda_{v}\tau_{z}+t(\cos k_{y}a+2\cos\frac{\sqrt{3}k_{x}a}{2}\cos\frac{k_{y}a}{2})\tau_{x}\nonumber\\
&&+t(\sin k_{y}a-2\cos\frac{\sqrt{3}k_{x}a}{2}\sin\frac{k_{y}a}{2})\tau_{y}-\lambda_{\rm R}
(\cos \frac{\sqrt{3}k_{x}a}{2}\sin\frac{k_{y}a}{2}+\sin k_{y}a)\tau_{x}s_{x}\nonumber\\
&&+\sqrt{3}\lambda_{\rm R}\sin \frac{\sqrt{3}k_{x}a}{2}
\cos\frac{k_{y}a}{2}\tau_{x}s_{y}+\lambda_{\rm R}
(\cos \frac{\sqrt{3}k_{x}a}{2}\cos\frac{k_{y}a}{2}-\cos k_{y}a)\tau_{y}s_{x}\nonumber\\
&&+\sqrt{3}\lambda_{\rm R}\sin \frac{\sqrt{3}k_{x}a}{2}
\sin\frac{k_{y}a}{2}\tau_{y}s_{y}.
\end{eqnarray}
This rank-4 matrix can not be decomposed as the direct sum of two rank-2 matrices any more, so we
have to use the general formula in Eq.(\ref{sgeneral}) and calculate the shift current tensor numerically.
Meanwhile, as the Rashba spin-orbit coupling does not break
the mirror symmetry about the $k_{x}=0$ plane, $\sigma^{xxx}$ and $\sigma^{xyy}$ still vanish identically.

\begin{figure}
\subfigure{\includegraphics[width=7cm, height=5cm]{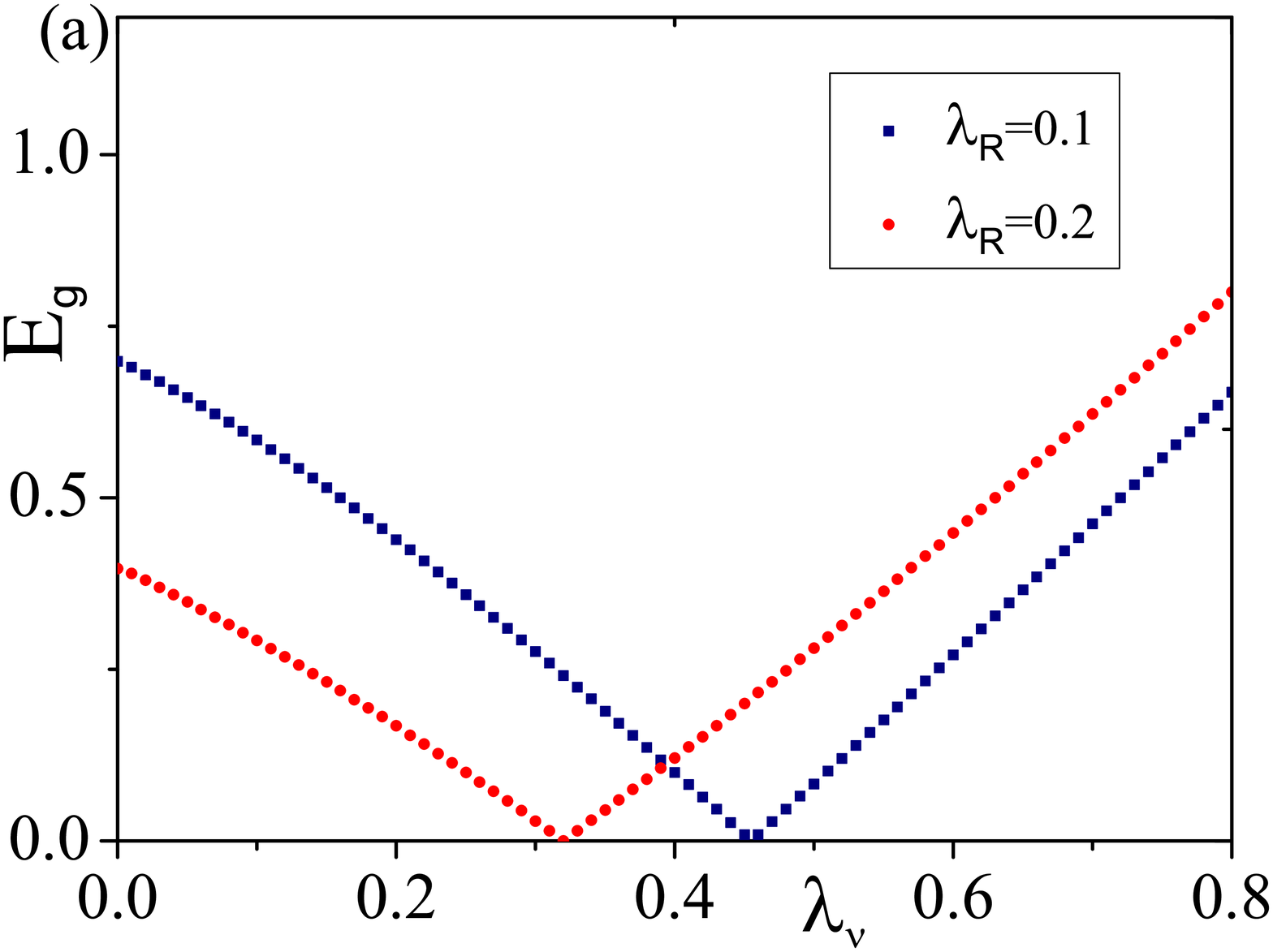}}
\subfigure{\includegraphics[width=7cm, height=5cm]{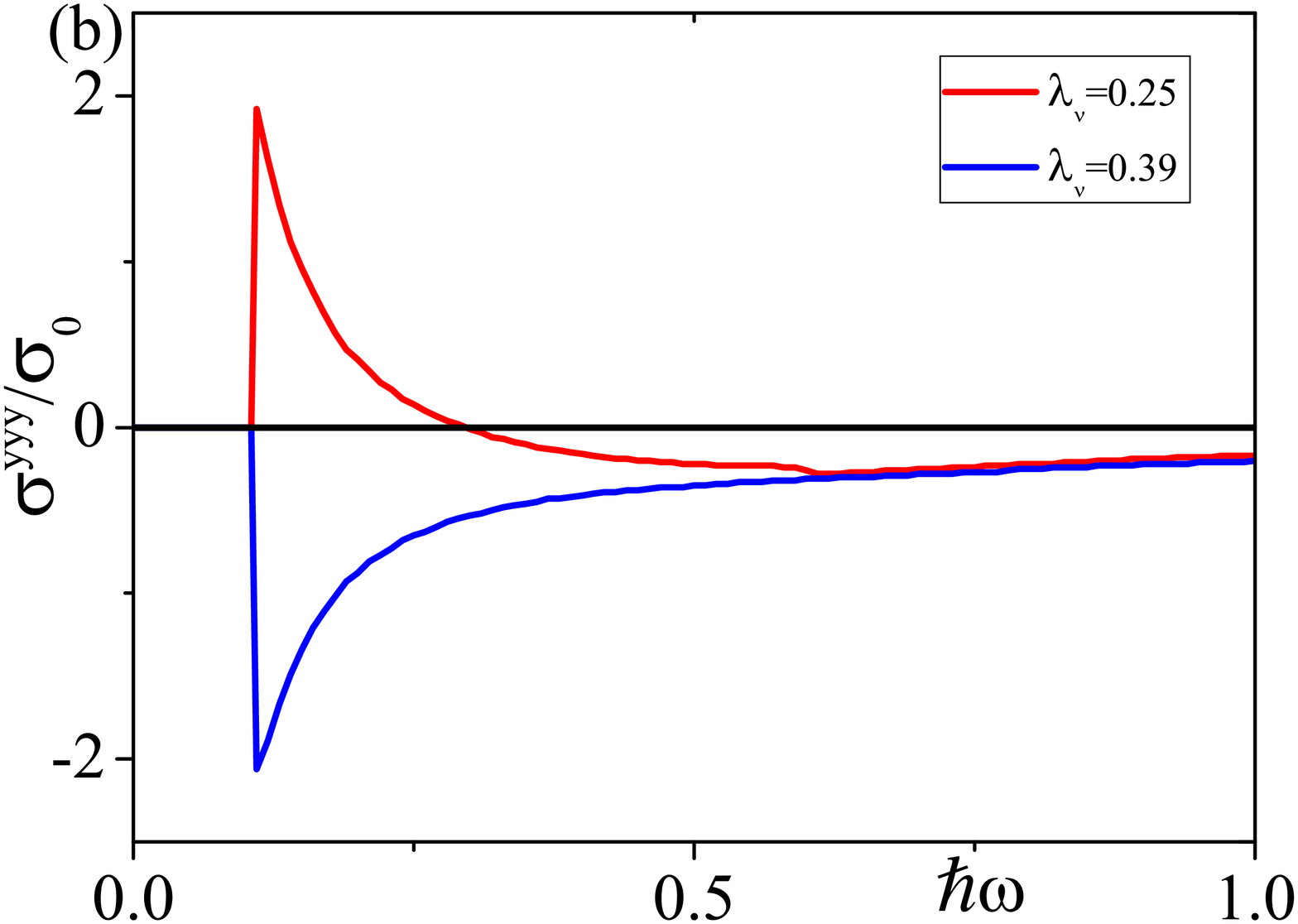}}
\caption{ Common parameters $t=1$, $\lambda_{\rm so}=\sqrt{3}/18$. (a) The variation of band gap with $\lambda_{v}$ for
fixed Rashba spin-orbit coupling.
The gap closing points ($E_{g}=0$) correspond to the critical points of TPTs. (b) Parameters $\mu=0$, $T=0$, $\lambda_{\rm R}=0.2$
and $\sigma_{0}=e^{3}/\hbar$.
According to the band gap evolution in (a), the TPT takes place at $\lambda_{v}=0.32$. Clearly, the band-edge shift current tensor
reverses its sign across the TPT, confirming that the sign-reversal behavior of band-edge shift current
tensor holds no matter whether the spin is conserved or not. }  \label{four-band}
\end{figure}

The numerical results are presented in Fig.\ref{four-band}. Fig.\ref{four-band}(a) shows that the introduction of  Rashba spin-orbit coupling will change
the position of critical point.  Fig.\ref{four-band}(b) demonstrates that the sign-reversal behavior of band-edge shift current holds
even when the spin conservation is broken by the Rashba spin-orbit coupling. Because we find that $\sigma^{yyy}\simeq-\sigma^{yxx}$ still holds,
here only $\sigma^{yyy}$ is presented.

\end{document}